\documentstyle[12pt]{article}
\topmargin - 0.8cm
\begin{document}
\setlength{\unitlength}{1mm}
{\hfill   December 1995 }

{\hfill    WATPHYS-TH-95-11 }

{\hfill    gr-qc/9512047 } \vspace*{2cm} \\
\begin{center}
{\Large\bf
Non-minimal coupling, boundary terms and renormalization of the Einstein-Hilbert action
and black hole entropy}
\end{center}
\begin{center}
{\large\bf A.O.~Barvinsky$^{1 \ast}$ and  S.N.~Solodukhin$^{2 \ast\ast}$}
\end{center}
\begin{center}
{\it $^1$ Theory Department of Lebedev Physics Institute and Lebedev
Research Center in Physics, Leninskii Prospect 53, Moscow 117924, Russia}
\end{center}
\begin{center}
{\it $^{2}$ Department of Physics, University of Waterloo, Waterloo, Ontario N2L 3G1, Canada  \\
and Bogoliubov Laboratory of Theoretical Physics, Joint Institute for
Nuclear Research, Dubna 141980, Russia}
\end{center}
\vspace*{2cm}
\begin{abstract}
A consistent variational procedure applied to the gravitational action
requires according to Gibbons and Hawking a certain balance between the
volume and boundary parts of the action. We consider the problem of preserving
this balance in the quantum effective action for the matter non-minimally
coupled to metric. It is shown that one has to add a special boundary term
to the matter action analogous to the Gibbons-Hawking one. This boundary term
modifies the one-loop quantum corrections to give a correct balance for the
effective action as well. This means that the boundary UV divergences do
not require independent renormalization and are automatically renormalized
simultaneously with their volume part. This result is derived for arbitrary
non-minimally coupled matter. The example of 2D Maxwell field is considered
in much detail. The relevance of the results obtained to the problem of the
renormalization of the black hole entropy is discussed.
\end{abstract}
\begin{center}
{\it PACS number(s): 04.60.+n, 12.25.+e, 97.60.Lf, 11.10.Gh}
\end{center}
\vskip 1cm
\noindent $^{\ast}$ e-mail: barvin@rq2.fian.msk.su \\
\noindent $^{\ast \ast}$ e-mail: sergey@avatar.uwaterloo.ca
\newpage
\section{Introduction}
\setcounter{equation}0
It was realised by Gibbons and Hawking \cite{GH} that a well-defined
variational procedure for the Einstein-Hilbert (EH) action
\begin{equation}
W_{EH}=-{1 \over 16\pi G}\int_M R
\label{1}
\end{equation}
on manifold  $M$ with boundary $\partial M$ requires fixing the metric
but not its normal derivative on the boundary. Therefore, they added to the
volume action
(\ref{1}) the boundary term the role of which is to compensate variations
of the normal derivatives of the metric  on $\partial M$ that come from the
variation of (\ref{1}) after integrating by parts. The resulting action takes
the form
\begin{equation}
W_{EH}=-{1 \over 16\pi G}\left( \int_M R+2\int_{\partial M} k \right)
\label{2}
\end{equation}
where $k=g^{\mu\nu} k_{\mu\nu}~~(k_{\mu\nu}={1\over 2}(\nabla_\mu n_\nu
+\nabla_\nu n_\mu ))$ is the extrinsic curvature of the boundary, $n^\mu$ is
outward pointing normal vector to $\partial M$. The variation procedure being
applied to (\ref{2}) is now consistent.

\bigskip

On the other hand, it is well known that the term like (\ref{1}) is typically
induced  by quantum corrections. And it was even proposed to treat the EH as
being completely induced by quantum matter fields interacting with background
classical gravitational field \cite{Sakharov}. However, it is natural to ask
whether the boundary term in EH (\ref{2}) can also be induced in such a way
that the ``correct'' balance between the volume and boundary parts of
(\ref{2}) is preserved. The related but somewhat more general question is
that does the quantum effective action, obtained by integrating out quantum
matter fields and being functional of background metric, have automatically
such a form to get the consistent variational procedure?

\bigskip

The  technique of the heat kernel expansion \cite{heat}
\begin{eqnarray}
&&W_{eff}={1 \over 2} \ln {\rm det} (-\Delta )=-{1 \over 2}
\int_{\epsilon^2}^{\infty} {ds \over s}  Tr  {K}_{M} (s),
\nonumber \\
&&{K}_{M} (s)=e^{s \Delta}={1 \over (4\pi s)^{d \over
2}}
\sum_{n}^{}{} {a}_n s^n, \ \ s \rightarrow 0
\label{3}
\end{eqnarray}
gives us the useful tool to investigate the problem. The EH like term in the
quantum effective action $W_{eff}$  appears in the first coefficient $a_1$ of
the expansion for
the matter field operator $\Delta$. For example, for the scalar field
minimally coupled to the gravitational field
\begin{equation}
W_{sc}={1 \over 2} \int_M \phi (-\Delta_0 )\phi~~ ,
\label{4}
\end{equation}
where $\Delta_0=\nabla_\alpha \nabla^\alpha$,
we get the promising result that
$$
a_1(\Delta_0)={1 \over 6} \left(\int_{M}^{}{R}+2\int_{\partial M}^{}k \right).
$$
However, already the non-minimally coupled scalar field
\begin{equation}
W^\xi_{sc}={1 \over 2} \int_M \phi (-\Delta^\xi_0 )\phi~~,
\label{6}
\end{equation}
where $\Delta^\xi_0=\Delta_0-\xi R$,
and vector field  described by the  action
\begin{equation}
W_{vec}={1 \over 2} \int_M A^\mu (-\Delta_{1})_{ \mu\nu} A^\nu ~~,
\label{7}
\end{equation}
where $(\Delta_{1})_{ \mu\nu}=g_{\mu\nu} \nabla^\alpha\nabla_\alpha-R_{\mu\nu}$
is the Beltrami-Laplace operator for one-forms, break this idyll:
\begin{eqnarray}
&&a_1(\Delta^\xi_0)=({1 \over 6}-\xi ) \int_{M}^{}{R}+{1\over 3}\int_{\partial M}^{}k~~; \nonumber \\
&&a_1 (\Delta_1)=({d \over 6}-1)\int_M R +{d \over 3} \int_{\partial M} k~~,
\label{8}
\end{eqnarray}
where $d=dim M$.

The reason for this obviously lies in the non-minimality of the coupling with
metric in (\ref{6}), (\ref{7}). Namely, the curvature tensors enter directly
the matter action (\ref{6})-(\ref{7}). Therefore,  if we calculate the
variation of the action with respect to background metric we observe the same
problem as for (\ref{1}): there are some non-zero variations of the normal
derivatives of the metric on the boundary. We need to add some boundary term
like in (\ref{2}) in order to kill these variations. More generally, before
demanding the quantum effective action to have the consistent metric variation
we need to start with the classical matter action satisfying this requirement.
The metric variation of the action gives us the stress-energy tensor
$T_{\mu\nu}$ of the matter. So, the modification of the actions
(\ref{6}), (\ref{7}) by appropriate boundary term would give us a well-defined
$T_{\mu\nu}$ without any peculiarities on the boundary. It is easy to find
such a modification for the scalar case (\ref{6}):
\begin{equation}
W_{mat}={1 \over 2}
\int_{M} \phi
(-\Delta^\xi_0)\phi +\xi\int_{\partial M}k \phi^2~~.
\label{9}
\end{equation}
It should be noted that the boundary term in (\ref{9}) appears even in the
flat space ($R=0$) when $\partial M$ has a non-zero extrinsic curvature.
Similarly, it is well-known that $T_{\mu\nu}$ for matter described by
(\ref{9}) is modified by $\xi$-dependent terms even in flat spacetime.

Now, if we start with the action (\ref{9}) and quantize $\phi$ we could expect
that the resulting effective action possesses
the needed property of a balance between the volume and boundary terms like
in (\ref{2}).
Indeed, the boundary terms like that in (\ref{9}) modify the heat kernel
expansion in the way to get the correct balance.
This is the aim of this paper to demonstrate how this happens in the general case of fields of arbitrary spin.

\bigskip

The action (\ref{2}) is a particular form of a more general expression\footnote{It should be noted that adding to (\ref{10}) quadratic term $U^2_{\mu\nu\alpha\beta}$ we would obtain the first order form of the
 higher-derivative  $R^2$-theory of gravity.}
\begin{equation}
W=\int_M U^{\mu\nu\alpha\beta}R_{\mu\nu\alpha\beta}-4\int_{\partial M} U^{\mu\nu\alpha\beta}
n_{\mu}n_\beta k_{\nu\alpha}~~,
\label{10}
\end{equation}
where $U^{\mu\nu\alpha\beta}$ is arbitrary tensor not containing derivatives
of the metric.
Below we prove that (\ref{10}) has a well-defined metric variation.
What we are really going to demonstrate is that the part of the effective
action which is of the first order in curvature always takes the form
(\ref{10}) for some tensor $U^{\mu\nu\alpha\beta}$, the concrete form of which
depends on the concrete non-minimal matter. For known types of matter
$U^{\mu\nu\alpha\beta}$ is some combination of the metric $g_{\mu\nu}$ but
not of its derivatives. Then (\ref{10}) certainly reproduces the form
(\ref{2}).

\bigskip

The coefficient $a_1$ of the heat kernel expansion (\ref{3}) typically
represents the power $1/\epsilon^{d-2}$ ultraviolet divergence of the
effective action. The fact that the structure of the divergent term repeats
the form (\ref{2}) is significant. This means that we do not need a special
renormalization for removing the UV divergences on the boundary as it could
happen for the theories (\ref{6})-(\ref{7}). Instead, all the divergences,
volume and at the boundary, are removed by only the renormalization
of the gravitational constant $G$ in (\ref{2}). 

One of the important applications of this result is the calculation and
renormalization of black hole entropy\cite{SU,FS,BFZ,Myers,S,LW,Kabat}.
For matter non-minimally coupled to gravity this calculation becomes
non-trivial. This was pointed in \cite{S} in the context of the conical approach 
to black hole entropy. The reason for this is that the Riemann tensors behave as 
distributions at the conical singularity. This fact gives rise to appearance of
contact interaction of the matter that is concentrated at the singularity
(that is horizon in the black hole case). As a result of this interaction,
one obtains the modification of the heat kernel expansion on the conical
space by terms defined on the singular subspace. Taking them into account
was shown to be important for the black hole entropy and allowed in \cite{S}
to extend the statement of simultaneous renormalization of  the entropy  and
effective action to the case of non-minimally coupled matter.
In the present paper we put this problem in another way and demonstrate
that the renormalization statement is closely related with
the ultraviolet renormalization of the volume and
surface part of the gravitational effective action. In this sight on the 
problem we are mostly close to the work
of Larsen and Wilczek \cite{LW} where it is argued that the universality of 
one renormalization for both the gravitational
constant $G$ and the entropy of a black hole follows from the low-energy
theorem -- the generic local structure of the low-energy effective action.
However, the work \cite{LW} does not reveal the concrete mechanism of
this phenomenon for the general case of non-minimal coupling, not to say
about its extension beyond the black hole entropy framework  (see discussion 
in Sect.6).
The purpose of this paper is to establish this mechanism and its
general validity at the one-loop order.

\bigskip

\section{The boundary term}
\setcounter{equation}0
We start with the action describing matter field of arbitrary spin which
is non-minimally coupled to gravity:
\begin{equation}
W_{mat}={1\over 2} \int_M \phi^A (-\Box_{AB}) \phi^B,
\label{2.1}
\end{equation}
with operator
\begin{equation}
\Box_{AB}=\eta_{AB} \nabla_\alpha \nabla^\alpha-U_{AB}^{\mu\nu\alpha\beta} R_{\mu\nu\alpha\beta},
\label{2.2}
\end{equation}
where $\{ \phi^A \}, ~~A=1,...,D$ is a section of the matter bundle over manifold $M$ with invariant
metric $\eta_{AB}$ and covariant derivative $\nabla_\alpha$. The tensor
$U_{AB}^{\mu\nu\alpha\beta}$ has symmetries of the Riemann tensor $R_{\mu\nu\alpha\beta}$ with respect to upper indexes. In principle it can be arbitrary tensor not containing metric derivatives. However, for known operators it is an combination of the metric $g_{\mu\nu}$ and $\eta_{AB}$.

In order to find the boundary term to be added to (\ref{2.1}), let us consider
a small vicinity of the boundary $\partial M$. There the manifold $M$ can be
represented as direct product $M=\partial M \otimes I$. Let the parameter $t$
label the hypersurfaces  of the foliation, the boundary $\partial M$ beeing
one of them. The outward pointing normal to the hypersurfaces is $n_\alpha=-N
\nabla_\alpha t$ where $N^2=(\nabla t)^2$ defines the lapse function $N$. The
hypersurface metric is $h_{\mu\nu}=g_{\mu\nu}-n_\mu n_\nu$. Then
the Gauss-Codazzi equation implies (see \cite{York}):
\begin{equation}
R_{\mu\nu\alpha\beta}=4n_{[\mu} {\cal L}_n k_{\nu ][\alpha}n_{\beta ]}+ ...
\label{2.3}
\end{equation}
where ${\cal L}_n$ is the Lie derivative along $n^\mu$. Since
$k_{\mu\nu}={1\over 2}{\cal L}_n h_{\mu\nu}$  the first term at r.h.s. of
(\ref{2.3}) is of the second order with respect to normal derivatives
${\cal L}_n{\cal L}_n h_{\mu\nu}$.
The (...) terms in (\ref{2.3}) are of lower order in the normal derivative.

Thus, under variation of the Riemann tensor in the expression
$$
\int_M U^{\mu\nu\alpha\beta}R_{\mu\nu\alpha\beta}
$$
only the first term in (\ref{2.3}) produces the second normal derivative  of the metric variation ${\cal L}_n{\cal L}_n \delta h_{\mu\nu}$ that after integration
by parts gives the variation on the boundary
$$
2\int_{\partial M}
U^{\mu\nu\alpha\beta}n_\mu n_\beta {\cal L}_n \delta h_{\nu\alpha}~~.
$$
This variation can be canceled if we add the  boundary term as follows
$$
\int_M U^{\mu\nu\alpha\beta}R_{\mu\nu\alpha\beta}-4\int_{\partial M} U^{\mu\nu\alpha\beta}
n_{\mu}n_\beta k_{\nu\alpha}~~.
$$
This is exactly the form announced in (\ref{10}).

Applying this result to (\ref{2.1})-(\ref{2.2}) we obtain the action with the boundary term
\begin{equation}
W_{mat}={1\over 2} \int_M \phi^A (-\Box_{AB}) \phi^B-2\int_{\partial M} \phi^A\phi^B U^{\mu\nu\alpha\beta}_{AB} n_\mu n_\beta k_{\nu\alpha}~~,
\label{2.4}
\end{equation}
which is our starting point for the quantization and derivation of the heat
kernel expansion.

\bigskip

\section{The heat kernel expansion}
\setcounter{equation}0
Considering the path integral over fields $\phi^A$ the dynamics of which is described by the action (\ref{2.4}) we get
\begin{eqnarray}
&&Z=\int [{\cal D} \phi ] e^{-W_{mat} } \nonumber \\
&&=\int [{\cal D} \phi ]~~e^{\int_{\partial M}\phi^A \phi^B V_{AB}}~~~
e^{-{1\over 2}\int_M \phi^A(-\Box_{AB})\phi^B}
\label{3.1}
\end{eqnarray}
where we denote $V_{AB}=2U^{\mu\nu\alpha\beta}_{AB} n_\mu n_\beta k_{\nu\alpha}$.
Taking the first term in (\ref{3.1}) perturbatively we can formally expand it
in powers of $V_{AB}$. In the leading order we get
\begin{equation}
Z=\bar{Z}\,\left(1+\big<\int_{\partial M} \phi^A \phi^B V_{AB}\big>_{\bar{Z}}+
\big<O(V_{AB}^2)\big> \right)~~,
\label{3.2}
\end{equation}
where the average $\big<~~ \big>_{\bar{Z}}$ is taken with respect to measure defined by functional integral
\begin{equation}
\bar{Z}=\int [{\cal D} \phi ] e^{-{1\over 2}\int_M \phi^A(-\Box_{AB})\phi^B}
\label{3.3}
\end{equation}
without boundary term.
Equivalently, we can write for the effective action $W_{eff}=-\ln Z$ ($\bar{W}_{eff}=-\ln
\bar{Z}$):
\begin{equation}
W_{eff}= \bar{W}_{eff} +\big< \int_{\partial M} \phi^A \phi^B V_{AB}+
O(V_{AB}^2)\, \big>_{\bar{Z}}~~.
\label{3.4}
\end{equation}
In the case when the boundary term is not included we have a standard
heat kernel expansion:
\begin{eqnarray}
&&\bar{W}_{eff}={1 \over 2} \ln {\rm det} (-\Box_{AB} )=-{1 \over 2}
\int_{\epsilon^2}^{\infty} {ds \over s}  Tr  \bar{K}_{M} (s),
\nonumber \\
&&\big<\phi^A (x) \phi^B (x')\big>_{\bar{Z}}= \int_{\epsilon^2}^\infty ds \bar{K}^{AB}_M (x,x',s), \nonumber \\
&&\bar{K}^{AB}_{M} (s)=e^{s \Box_{AB}}={1 \over (4\pi s)^{d \over
2}}
\sum_{n}^{}{} \bar{a}^{AB}_n s^n, \ \ s \rightarrow 0
\label{3.5}
\end{eqnarray}
where $n$ in the sum runs $0,~1/2,~1,~3/2, ...$ .
For manifold with boundary one typically imposes some boundary condition on
the quantum field $\phi^A$: $\phi^A|_{\partial M}=0$ for the Dirichlet
condition and ${\cal L}_n \phi^A|_{\partial M}=0$ for
the Neumann one. Correspondingly, this condition is imposed on the heat kernel
$\bar{K}_M(x,x',s)$ when one of the points $x$ or $x'$ lies on the boundary.
Therefore, one could naively expect that the second term in (\ref{3.4}) is
zero for the Dirichlet condition. However, this does not happen since
the limit of the coincident points is considered, which is rather peculiar
(see the derivation based on the method of images in \cite{McKean}). In
particular, we have
\begin{equation}
\bar{a}_{0 AB}(x,x)=\eta_{AB}
\label{*}
\end{equation}
even if $x$ lies at $\partial M$, that is consequence of the other condition
$$
\bar{K}_M(x,x',0)=\delta (x,x')
$$
imposed at $s=0$.
The first few coefficients of the expansion for the operator (\ref{2.2}) can
be found from known results (see for example \cite{McKean,Gilky}).
In particular,  the trace of $\bar{a}_{1AB}(x,x')$ is given by
\begin{equation}
\bar{a}_1 (\Box ) = \int_M \left({D \over 6}R -\eta^{AB}U_{AB}^
{\mu\nu\alpha\beta} R_{\mu\nu\alpha\beta} \right) +
{D \over 3} \int_{\partial M} k.
\label{3.6}
\end{equation}
Note, that (\ref{*}) and (\ref{3.6}) are the same for both Dirichlet and
Neumann conditions. Inserting  (\ref{3.5}) into (\ref{3.4}) we obtain
\begin{eqnarray}
&&W_{eff}=-{1 \over 2}
\int_{\epsilon^2}^{\infty} {ds \over s}  Tr  {K}_{M} (s),
\nonumber \\
&&Tr K_M (s)=Tr \bar{K}_M (s) +2s Tr_{\partial M} \left( V \bar{K}_M (s)
\right)+O(V^2)~~,
\label{3.8}
\end{eqnarray}
where the $x$-integration in $Tr_{\partial M}$ is taken over only the
boundary $\partial M$.

 From (\ref{3.8}) we have the following expression for the heat kernel
$K_{M}(s)$:
\begin{eqnarray}
&&Tr K_{M} (s)= {1 \over (4\pi s)^{d/2} } \sum_{n}a_n s^n,
\nonumber \\
&&a_n=\int_{M} \bar{a}_n (x,s)+2\,\int_{\partial M} Tr \left( V
\bar{a}_{n-1}(x,s) \right)+O(V^2).
\label{3.9}
\end{eqnarray}
Note that typically every new power of $V$ in the perturbation theory for the
heat kernel brings at least one extra power of $s$ (see the derivation in
\cite{heat}), whence $O(V^2)$ term here contributes to $a_n$ starting with
$a_2$ and does not affect the calculation of $a_1$. Thus, for the $a_1$
coefficient we obtain, taking into account (\ref{*}) and (\ref{3.6}) that
\begin{equation}
a_1(\Box )={D\over 6} \left( \int_M R+2\int_{\partial M}k \right)-
\left( \int_M \eta^{AB} U^{\mu\nu\alpha\beta}_{AB}R_{\mu\nu\alpha\beta}-4\int_{\partial M}\eta^{AB} U^{\mu\nu\alpha\beta}_{AB}
n_{\mu}n_\beta k_{\nu\alpha} \right)
\label{3.10}
\end{equation}
The Eq.(\ref{3.10}) is our main result. It shows that the linear in the
curvature term of the effective action for non-minimal matter fields
indeed repeats the EH form (\ref{2}) (or, more generally, the form
(\ref{10})), if we include the boundary term as in (\ref{2.4}).
Though there exists possibility to consider arbitrary tensor $\eta^{AB} U^{\mu\nu\alpha\beta}_{AB}$,
not necessarily related to metric, for known types of matter it is the combination of the metric tensor.
Then the second term in (\ref{3.10}) repeats with some overall coefficient the form of the first term.
In the particular cases of non-minimal matter considered in the Introduction,
we have $D=1$
for the scalar  (\ref{6}) and $\eta_{AB}\equiv g_{\mu\nu}$, $D=d\equiv dim M$ for vector  (\ref{7}) and in the both cases (up to factor $\xi$ for scalar ) we get that $\eta^{AB}U^{\mu\nu\alpha\beta}_{AB}={1\over 2}
(g^{\mu\alpha}g^{\nu\beta}-g^{\mu\beta}g^{\nu\alpha} )$. Then, the
corresponding coefficients $a_1$ calculated  according to (\ref{3.10}) read
\begin{equation}
a_1(\Delta^\xi_0)=({1\over 6}-\xi ) \left(\int_M R+2\int_{\partial M}k \right)
\label{3.11}
\end{equation}
\begin{equation}
a_1(\Delta_1)=({d\over 6}-1 ) \left(\int_M R+2\int_{\partial M}k \right) ~~.
\label{3.12}
\end{equation}
Eq. (\ref{3.9}) allows calculate other coefficients of the heat kernel expansion for both
integer and half-integer $n$. We are not doing this here.

The functional integral (\ref{3.1}) can be written as an average of the
boundary operator $\big<\hat{V}[{\partial M}]\big>=\big<e^{\int_{\partial M}
\phi^A \phi^B V_{AB}}\big>$. It is worth noting that this operator is similar to
other objects that appeared earlier in field theory models: Wilson
loop $\big<P e^{\int_C A_\mu dx^\mu}\big>$ in the theory
of non-Abelian gauge fields and vertex operator $V=e^{\imath p X(z)}$
describing the contact interaction in string theory. Following this analogy,
we may interpret our boundary operator as describing some (contact)
interaction at the boundary.

\bigskip

\section{2D Maxwell field}
\setcounter{equation}0
As a simple application of the results obtained let us consider Maxwell field
on two-dimensional manifold with boundary.
 We define the
partition function for the Maxwell field, including the contribution of
ghosts,  as follows
\begin{equation}
Z=[{\rm det}' \Delta_1 ]^{-1/2} {\rm det}' \Delta_0
\label{A1}
\end{equation}
where ${\rm det}'$ is calculated only on non-zero  modes of operators;
$\Delta_k=(d \delta+\delta d)_{(k)} $ is the Beltrami-Laplace operator
acting on $k-$forms.

In two dimensions we have a remarkable property for closed manifold
that the set of non-zero eigenvalues of
$\Delta_1$  is given by a union non-zero eigenvalues of  $\Delta_0$ and
$\Delta_2$. This is simply a consequence of the cohomological algebra of the
operators $\Delta_k$, $\delta_k$ and $d_k$ (see for example \cite{Hodge}).
Moreover, due to Hodge dualization in two dimensions the eigenvalues of
operators $\Delta_0$ and $\Delta_2$ are the same. Hence we have that
${\rm det}' \Delta_1= {\rm det}'\Delta_0 {\rm det}' \Delta_2= ({\rm det}'
\Delta_0)^2$ and, therefore, (\ref{A1}) is trivial, $Z=1$. In particular,
there is not any ultraviolet divergence in $\ln Z$.
The cohomological arguments can be applied to the open manifold as well
and one could expect the same result.

To proceed with the heat kernel, it is worth noting that
 for an arbitrary elliptic operator $A$ the formula (\ref{3}) is modified
due to zero modes
\begin{equation}
\ln {\rm det}'A =-\int\limits_{\varepsilon^2}^{\infty}{dt \over t}
Tr\,[ e^{-tA}-P(A)]
\label{A2}
\end{equation}
where $P(A)$ is a projector onto the subspace of zero modes of $A$;
$Tr P(A)=N(A)$ -- the number of zero modes of $A$. In $d$ dimensions the
zero modes contribute to the coefficient $a_{d/2}$ in the heat kernel
expansion.

For the divergent part of (\ref{A1}) in two dimensions we have
\begin{equation}
(\ln Z)_{div}=[{1 \over 2\pi}[{1 \over 2} a_1 (\Delta_1)-a_1 (\Delta_0)]+
 [2N_0-N_1]] \ln {L \over \varepsilon}
\label{A15}
\end{equation}
where $N_k=N(\Delta_k)$.  For a closed 2D manifold the combination of
numbers ($2N_0-N_1$) in
(\ref{A15}) is the Euler number\footnote{The general definition of the Euler number
in two dimensions reads: $\chi (M)=N_0-N_1+N_2$. However, for a 2D closed manifold
we have by Hodge duality that $N_2=N_0$ and hence
$\chi (M)=2N_0-N_1$.}  $\chi (M)={1 \over 4\pi}( \int_{M}^{}R
+2\int_{\partial M}^{}k)$ of 2D manifold.
Using a standard  (\ref{8}) or modified (\ref{3.12}) expression for the
coefficients (for a closed manifold they are the same) we obtain that
 (\ref{A15}) indeed vanishes.

For an open 2D manifold the situation is more complicated. First of all, we 
have to
impose some boundary conditions on $p$-forms eigen vectors in the partition 
function
(\ref{A1}). The possible choice is  to put Dirichlet condition for
zero-forms and the generalized Dirichlet condition for the one-form 
$A=A_\mu dx^\mu$: $A^\mu \epsilon_{\mu\nu} n^\nu |_{\partial M}=0$, 
where $n^\mu$ is a unit vector normal to the boundary.  For a disk with 
polar coordinates $(r,\phi )$ the later conditions means $A_\phi=0$ on the 
boundary. The important observation now is that
the combination $(2N_0-N_1)$ is no longer equal to the Euler number of the 
manifold. Indeed, for boundary conditions as above for a disk we have 
$N_0=N_1=0$ while the Euler number of a disk is $\chi=1$. But the 
combination $(2N_0-N_1)$ is still a metric independent quantity and its 
metric variation vanishes.  Then, using the standard coefficients 
(\ref{8}) in (\ref{A15}) for open manifold we obtain  the disbalance between
volume $\int_M R$ and boundary $\int_{\partial M} k$ parts. The balance 
certainly restores if we take
into account the boundary operator according to (\ref{3.1}) when calculating
the quantity $\ln {\rm det}' (-\Delta_1)$ in (\ref{A1}).
This is easily checked by inserting the  expression (\ref{3.12}) for
$a_1(\Delta_1)$ in (\ref{A15}) instead of the standard one. We then obtain
for an open manifold
\begin{equation}
(\ln Z)_{div}= [-\chi (M)+
 (2N_0-N_1)] \ln {L \over \varepsilon}~~.
\label{A16}
\end{equation}
The metric variation of (\ref{A16}) is well-defined. However, it
does not take exactly the form (\ref{2}) due to the contribution of zero 
modes.
Note, that the form of the heat kernel coefficients is universal being 
independent
of the dimension of the manifold. Otherwise, the contribution of 
zero-modes in
(\ref{A2}) is essentially dependent on spacetime dimensionality.
Therefore, the structure of (\ref{A16}) is very special and appears 
only in two dimensions.
The UV divergence (\ref{A16}) is similar to that of obtained by Kabat 
\cite{Kabat}
though in his calculation he neglected the role of terms on the boundary
$\partial M$ and contribution of zero modes.

\bigskip

\section{Renormalization of black hole entropy}
\setcounter{equation}0
The calculation of the heat kernel in Sect.3 is very similar to the
calculation of the (divergent) quantum correction to the black hole entropy
in \cite{S}.  This fact is certainly not occasional. In order to demonstrate
that our results are relevant to the black hole entropy we note that it is
a boundary term in (\ref{2}), or in a more general expression (\ref{10}),
that is responsible for the entropy. There are different ways to show this,
we follow ref.\cite{Teitelboim}.

Consider the Euclidean black hole instanton with metric
\begin{equation}
ds^2=g(\rho)d\tau^2+d\rho^2+r^2(\rho) d\Omega^2~~,
\label{5.1}
\end{equation}
where $d\Omega^2=\gamma_{ab}(\theta )d\theta^a d\theta^b$ is metric of
 $(d-2)$-shpere of unity
radius; the period of $\tau$ in (\ref{5.1}) is chosen to remove the singularity at the horizon
surface $\Sigma$ defined by the equation $g(\rho_\Sigma)=0$. We take
the coordinate $\rho$ such that $\rho_\Sigma=0$, then
$g(\rho)=\rho^2/\beta_H^2+ O(\rho^4)$ and $r^2(\rho )=r^2_\Sigma +O(\rho^2)$. Take a small ball $B_\delta$ of
radius $\delta ~~(0 \leq \rho \leq \delta)$
surrounding the surface $\Sigma$.
Vector normal to $\partial B_\delta$ has only one non-zero component $n_\rho =1$.
Then the components of the extrinsic curvature of $\partial B_\delta$
read $k_{\mu\nu}={1\over 2} \partial_\rho g_{\mu\nu}$.
 When $B_\delta$ shrinks
($\delta \rightarrow 0$) the boundary $\partial B_\delta$ goes to $\Sigma$.

Take the gravitational action in the form (\ref{10}) on $B_\delta$
\begin{equation}
W[B_\delta ]=\int_{B_\delta} U^{\mu\nu\alpha\beta}R_{\mu\nu\alpha\beta}-4\int_{\partial B_\delta } U^{\mu\nu\alpha\beta}
n_{\mu}n_\beta k_{\nu\alpha}~~.
\label{5.2}
\end{equation}
The volume term in (\ref{5.2}) then vanishes when $\delta \rightarrow 0$
while the boundary term gives
the integral over the surface $\Sigma$
\begin{equation}
W[B_{\delta \rightarrow 0} ]=-4\pi \int_\Sigma U^{\mu\nu\alpha\beta}n^i_\mu n^i_\alpha n^j_\nu n^j_\beta~~,
\label{5.3}
\end{equation}
where $\{ n^i \}$ is a pair of vectors normal to $\Sigma$ (the only non-zero
components are $n^1_\rho=1,~n^2_\tau={\rho \over \beta_H}$).
In fact, the boundary term in (\ref{5.2}) produces combination of the type
$ \sqrt{\gamma} \rho (U^{\rho\tau\tau\rho} k_{\tau\tau}+U^{\rho ab\rho} k_{ab})$,
where for small $\rho$ we have for the extrinsic curvature: $k_{\tau\tau}={\rho \over \beta^2_H}+O(\rho^3);~~k_{ab}={1\over 2} \gamma_{ab} \partial_\rho r^2$. Only component $U^{\rho\tau\tau\rho}$ is assumed to be divergent (as $1\over \rho^2$) in the center of the polar coordinates $(\rho , \tau )$ while components $U^{\rho ab\rho}$ lie in the orthogonal space and remain finite for $\rho=0$. Therefore, taking limit $\rho=\delta \rightarrow 0$ we obtain the result (\ref{5.3}).
 The expression (\ref{5.3})
coincides exactly with (minus) the black hole entropy
calculated by a number of other methods (see recent paper \cite{Brown}
where the different methods are compared). So, the entropy of a black hole
(at least for the theories linear in curvature) can be treated as a
gravitational action
\begin{equation}
S_{BH}=-W[B_{\delta \rightarrow 0} ]
\label{5.4}
\end{equation}
defined for the infinitesimal ball $B_{\delta \rightarrow 0}$ surrounding
the horizon surface $\Sigma$. Note, that above calculation is essentially
off-shell that makes it similar to the calculation within the conical method
of \cite{Teitelboim,FS}. Other remark is that this calculation can be also applied
to a higher-derivative theory of gravity if the later preliminary re-expressed
in the first order form (see \cite{Brown}).

One can see now that there is a strong correlation between the volume and
boundary terms in the classical gravitational action (\ref{2})
(or (\ref{10})) due to the necessity of a consistent variational procedure.
This correlation is preserved, as we have shown, in the divergent part of
the quantum effective action. Therefore, we need to renormalize only the
gravitational constant $G$ to remove both the volume and boundary parts of
the UV divergences. The boundary divergent term of the effective action
(see (\ref{3.8})-(\ref{3.10})) by the same line of reasoning as in
(\ref{5.3})-(\ref{5.4}) gives the divergence of the entropy
\footnote{We do not consider here the logarithmic ($\ln \epsilon$)
divergence (for $d>2$) of the entropy originating from $R^2$-terms
in the effective action \cite{FS}.}
\begin{equation}
S_q={1\over \epsilon^{d-2}}{1\over (d-2)}{1\over (4\pi)^{d-2\over 2}}
\left( {D\over 6} \int_\Sigma 1 +
\int_\Sigma \eta^{AB} U_{AB}^{\mu\nu\alpha\beta}n^i_\mu n^i_\alpha n^j_\nu n^j_\beta \right)
\label{5.5}
\end{equation}
due to the non-minimally coupled matter of the general type (\ref{2.4}),
which is a quantum addition to its classical counterpart
(\ref{5.3})-(\ref{5.4}). Since the entropy is related to the boundary term
in the action, we obtain from our consideration a simple proof of the
statement that the black hole entropy is automatically renormalized by the
same procedure as the effective action. This is just the consequence of
(\ref{5.4}) and of the balance between the volume and boundary parts
established above for the quantum effective action.

For the scalar (\ref{6}) and one-form  (\ref{7}) matter we get the quantum
entropy correspondingly as follows:
\begin{equation}
S^{sc}_q={1\over \epsilon^{d-2}}{1\over (d-2)}{1\over (4\pi)^{d-2\over 2}}
\left( {1\over 6}-\xi\right) \int_\Sigma 1
\label{5.6}
\end{equation}
\begin{equation}
S^{vec}_q={1\over \epsilon^{d-2}}{1\over (d-2)}{1\over (4\pi)^{d-2\over 2}}
\left( {d\over 6}-1 \right) \int_\Sigma 1
\label{5.7}
\end{equation}
The result (\ref{5.6}) coincides with that previously
obtained in \cite{S}.

The results of Section 4 can be used to study the renormalization of 
the Maxwell field in two dimensions.
The use of boundary conditions considered in Sect.4 allows one to neglect 
the contribution of the zero modes in (\ref{A16}). Then the Maxwell field
in two dimensions gives rise to a constant (UV-divergent) contribution
to the entropy which is renormalized in the same manner as 2D (effective)
gravitational coupling. The non-zero result for the entropy of Maxwell fields
seems puzzling (see \cite{Kabat}) in view of the absence of their propagating
degrees of freedom in two dimensions. Its statistical entropy,
therefore, is expected to be zero \cite{Kabat}. A possible resolution of 
this puzzle is that our method of calculation yields the thermodynamical 
entropy which may differ from the statistical one by the constant 
independent of spacetime geometry. The same happens with 2D topological 
gravity described by the action
$$
W_{top}=\int_M R +2\int_{\partial M} k.
$$
This model has a constant contribution to entropy, while $W_{top}$ does not 
describe any dynamical degrees of freedom. In view of this, the fact that 
the Maxwell field entropy is
just a (UV-divergent) constant, independent on the black hole geometry,
can be considered as a manifestation of trivial nature of this field in
two dimensions.

To make the correspondence with the conical method considered in
\cite{SU,FS,S} note that the effect of the conical singularity is
concentrated in the infinitesimal region near the singular surface
participating in the construction of (\ref{5.4}). On the other hand, we do
not concern here the higher curvature terms in the effective action.
It is not quite clear how to
generalize the considerations of this paper to include such terms.

Another problem to be mentioned is whether the same balance is true for 
other boundary conditions on the metric. Indeed,
instead of fixing the metric on the boundary (Dirichlet problem)
we could fix its normal derivative that changes the boundary term in the 
gravitational action. These questions in more detail will be considered elsewhere.

\bigskip

\section{Conclusion}
\setcounter{equation}0
The main question addressed in this paper is whether the Einstein-Hilbert
action
can be generated by quantum matter exactly in the form suggested by  Gibbons
and Hawking that possesses the consistent variation with respect to metric
subject to Dirichlet conditions for the metric coefficients of the
boundary.
We argue that in order to get this one has to start with a matter action the
metric
variation of which on manifold with a boundary is well defined. It is shown
that the action of matter non-minimally coupled with metric requires some
special boundary term analogous
to the Gibbons-Hawking one. We derive this term for arbitrary non-minimal
matter.
Then in the effective action of the quantized matter the term of the first
order in the curvature
is generated in the correct form.  In particular, this means that the
corresponding boundary UV divergences do
not require an independent renormalization and are automatically renormalized
simultaneously with their volume part linear in the curvature. 
We relate this fact to the problem  of the renormalization
of a black hole entropy and arrive at the same result as \cite{S}.
The similar conclusion was done by the authors of \cite{LW}.
We should emphasize, however, the essential difference in
methods and generality of results of the present paper from those of
\cite{LW}.

For the renormalization of the gravitational coupling constant and black hole
entropy the authors of \cite{LW} used two different methods adjusted
correspondingly to two different definitions of entropy: the
Gibbons-Hawking thermodynamic entropy and the so-called geometric one
(defined by differentiating the effective action with respect to the
deficit angle on the conical manifold). The first method is a standard local
Schwinger-DeWitt expansion and proper time regularization on the manifold
with smooth metric and sufficiently small curvature. The second one
uses the trick of decomposing the spacetime in the vicinity of a conical
singularity into a product of two spaces (two-dimensional and transversal,
angular, one) with a subsequent use of a powerful two-dimensional machinery.
Generically, the conical singularity technique  for 
 non-minimally coupled 
matter  was developed  in \cite{S}.
The first method allows the authors of \cite{LW} to establish the validity
of the main result only for a minimally coupled scalar field, which boils
down to the statement of eq.(1.5) in our Introduction, because they
do not use the correct form of the boundary action like in (1.10) and
(2.4) and do not include its contribution in the expansion (3.2)
and in the final result (3.10). Without this it is impossible to reach
a correct result for generic nonminimally coupled fields. So, as a remedy
for a general case the second method is used in \cite{LW} reproducing the
results of \cite{LW} for the particular case of $2+(D-2)$ decomposition near the
horizon. The authors emphasize a conceptual inequivalence of these
two methods and domains of their applicability, but reckognize a miraculous
coincidence of their results in case of a minimal scalar field. In our 
derivation we don't use the conical singularity method that has a disadvantage
of manipulating with singular operators.
To arrive at conclusions of a
common renormalization of the volume Einstein-Hilbert term and the
relevant surface term for arbitrary boundary (not necessarily related to
conical singularities) we have to use in all entirety a regular technique of
the present paper.

The last comment concerns the use of low-energy theorem arguments
in [13] for a better foundation of the main result. Certainly it seems
tempting to declare that the correctness of the variational problem
for a local low-energy effective action should guarantee the same
renormalization of its bulk and surface terms. In reality, however, such
arguments have only heuristic nature and should be supported by verifying
the quantitative mechanism of this phenomenon. There are examples, maybe
in the different context, when quantum corrections can potentially
violate the asymptotically flat boundary conditions given at the classical
level, and nontrivial intrinsic cancelation of dangerous terms should
be checked to maintain one and the same boundary value problem in
classical and quantum domains \cite{MV}. Another example is the same problem
as posed in the present paper but with Neumann boundary conditions for metric.
It was used for constructing the microcanonical gravitational ensemble
\cite{BY}, when instead of the 3-metric of the boundary the quasilocal energy
and momentum are being fixed there. At present, it is far from obvious for us
if and how the same conclusions will hold for this problem, although the
low-energy theorem argument would seem to be equally applicable.

The range of open issues related to this work can be further continued.
It is not clear whether this can be generalized to
terms of higher powers in curvature in the effective action. It seems
reasonable that starting with a matter action having a well defined metric
variation we would obtain under quantization the effective action
in the right form to have the same property. However, the general proof of
this, even in
the case of a minimal matter (\ref{4}), is still absent. Moreover, the
formulation
of the consistent variational problem for the full effective action
on spacetimes with boundaries is not yet clear due to its essentially nonlocal
nature.

\bigskip

\begin{center}
{\bf Acknowledgments}
\end{center}
We thank I.Avramidi, D.Kabat, R.Myers and D.Vassilevich for useful conversations.
One of the authors (A.O.B.) is grateful for the support of this work
provided by the Russian Foundation for Fundamental Research under
Grant 93-02-15594, International (Soros) Science Foundation and
Government of the Russian Federation Grant MQY300 and the European
Community Grant INTAS-93-493. Partly this work has been made also possible
due to the support by the Russian Research Project "Cosmomicrophysics".
Research  of S.N.S. is supported by NATO Fellowship and in part by the Natural
Sciences and Engineering Research Council of Canada.

\end{document}